\documentstyle[12pt]{article}
\begin{document}
\rightline{TUHE-9911}
\rightline{hep-ph/9901294}
\begin{center}
{\bf\large New experimental tests of sum rules\\
for charmed baryon masses}\\
\vspace{16pt}
Jerrold Franklin\\
{\it Department of Physics, Temple University,\\
Philadelphia, Pennsylvania 19122-6082}\\
V5030E@VM.TEMPLE.EDU\\
January 13, 1999
\end{center}
\begin{abstract}

New experimental measurements are used to test
model independent sum rules for charmed baryon masses.
Sum rules for medium-strong mass differences are found to be reasonably well
satisfied with increasing accuracy, and the new measurements permit an
improved prediction of $2778\pm 9$ MeV
for the mass of the $\Omega_c^{*0}$.
But an isospin breaking sum rule
for the $\Sigma_c$ mass splittings is still in significant disagreement posing a
serious problem for the quark model of charmed baryons.   Individual
$\Sigma_c$ mass splittings are investigated, using
the new CLEO measurement of the $\Xi_c^\prime$
mass splitting, but the accuracy is not yet sufficient
for a good test.
\end{abstract}
\vspace{.5in}

Model independent sum rules[1-3] were derived some time ago for heavy-quark
baryon masses using fairly minimal assumptions within the quark model.
The sum rules depend on standard quark model
assumptions, and an additional assumption that the interaction
energy of a pair of quarks in a particular spin state does not
depend on which baryon the pair of quarks is in (``baryon independence").
This is a somewhat weaker assumption than full SU(3) symmetry of the wave
function, which would require  the same spatial wave function for each octet
baryon,
and each individual wave function to be SU(3) symmetrized.
 Instead, we use wave functions with no SU(3) symmetry, as described in
Ref.\cite{sqm}.
The wave functions can also be different for different quarks.  For instance, a
u-s pair in the $\Sigma^+$ hyperon can have a different spatial wave function
than  a u-d pair in the proton,
but is assumed to have the same interaction energy as a u-s pair in the $\Xi^0$
hyperon.

In deriving the sum rules, no assumptions are made about the type of potential,
and no
internal symmetry beyond baryon independence is assumed.
The sum rules allow any amount of symmetry breaking in the interactions and
individual wave functions, but do rest on baryon independence for each
quark-quark interaction energy.
Several of the sum rules [Eqs.\ (4), (5), and (6) below] also rely on the
assumption that there
is no orbital angular momentum so that the three spin-$\frac{1}{2}$ quark spins
add directly
to spin-$\frac{1}{2}$ or spin-$\frac{3}{2}$.
More detailed discussion of the derivation of the sum rules is given in Refs.\
[1] and [4].

We have previously tested these sum rules in Refs.\ \cite{cb2} and \cite{cb3}
using early measurements of heavy-quark baryon masses.  Those tests showed
reasonable agreement within fairly large experimental errors for two sum rules
for medium-strong charmed baryon mass differences and for one sum rule for
bottom baryon mass differences.
But there was a relatively large, and worrisome, discrepancy for the isospin
breaking mass differences between the $\Sigma_c$ charge states.  Since those
tests, there have been a number of  new experiments[6-11] resulting in more
accurate and more reliable values for some of the charmed baryon masses used in
the sum rules.  In this paper we look at the effect on the sum rules of these
new experiments, especially the recent CLEO II measurement\cite{c98} of the
$\Xi_c^{\prime +}$ and $\Xi_c^{\prime 0}$ masses.

 The measured charmed baryon masses that will be used in the sum rules are
listed in
table I for the expected baryon assignments.
The $\Xi_c^{+}$ baryon and the  $\Xi_c^{\prime+}$ baryon are distinguished, in
the
quark model, by having different spin states for the u-s quark pair.  The
$\Xi_c^{+}$ is the spin-$\frac{1}{2}$ usc baryon having the
u-s  quarks in a spin zero state, and the $\Xi_c^{\prime+}$ has the u-s quarks
in a
spin one state.  A similar distinction is made for the d-s quark pair in the
$\Xi_c^{0}$ and $\Xi_c^{\prime0}$ charmed baryons.
The numerical values in Table I are given in terms of appropriate mass
differences when that corresponds to how the measurement was made.
Where new experiments have given more accurate numbers since our
 previous test of the sum rules, a star has been put after the reference.
 Masses for light quark (u,d,s) baryons are all taken from the Review of
Particle Physics\cite{pdg}.
\begin{table}
\centering
\begin{tabular}{llc}
Baryon & Mass (MeV) & Reference\\
\hline\hline
$\Lambda_c^+$ & $2284.9\pm .6$ &\cite{pdg}\\
$\Sigma_c^{++}$ & $\Lambda_c^+ +167.9\pm .2 $&\cite{pdg,ait,frab96}*\\
$\Sigma_c^{0}$ & $\Sigma_c^{++}-0.6\pm .2 $&\cite{pdg,ait}*\\
$\Sigma_c^{+}$ & $\Sigma_c^0 +1.4\pm .6 $&\cite{crawf}\\
$\Sigma_c^{*++}$ & $\Lambda_c^+ + 234.5\pm 1.4$ & \cite{brand}*\\
$\Sigma_c^{*0}$ & $\Lambda_c^+ + 232.6\pm 1.3$ & \cite{brand}*\\
$\Xi_c^{+}$ & 2465.6$\pm 1.4$ &\cite{pdg,edw}*\\
$\Xi_c^{0}$ & 2470.3$\pm 1.8$ &\cite{pdg}\\
$\Xi_c^{\prime+}$ & $\Xi_c^+ +107.8\pm 3.0$ &\cite{c98}*\\
$\Xi_c^{\prime0}$ & $\Xi_c^0 +107.0\pm 2.9$ &\cite{c98}*\\
$\Xi_c^{*+}$ & $\Xi_c^0 +174.3\pm 1.1$ &\cite{gib}*\\
$\Xi_c^{*0}$ & $\Xi_c^+ +178.2\pm 1.1$ &\cite{av95}\\
$\Omega_c^0$ & 2704$\pm$4 & \cite{pdg}\\
\hline
\end{tabular}
\caption{Charmed baryon masses used in the sum rules.}
\end{table}\newpage

The isospin breaking sum rule for the $\Sigma_c$ masses is\cite{cb2}
\begin{equation}
\Sigma^+ +\Sigma^-  -2\Sigma^0
 = \Sigma^{*+}+\Sigma^{*-}-2\Sigma^{*0}
 = \Sigma_c^{++}+\Sigma_c^0-2\Sigma_c^+,
\end{equation}
\vskip -.09in
\hspace{.61in}$(1.7\pm .2)$\hspace{.61in}$(2.6\pm 2.1)$\hspace{.77in}$(-2.2\pm
1.2)$\\
where we have written the experimental values in MeV below each equation.
There is reasonable agreement for the $\Sigma-\Sigma^*$ sum rule, as well as
for several other isospin breaking sum rules for light quark
baryons\cite{cb,sqm}.
But the $\Sigma_c$ isospin splitting combination is significantly different from
the other two combinations in Eq.\ (1).  As noted in Ref.\  [2], this
disagreement poses a serious problem because it is difficult to see how any
reasonable quark model of charmed baryons could lead to the relatively large
negative value for the $\Sigma_c$ combination in Eq.\ (1).  A large number of
specific quark model calculations\cite{theo} of charmed baryon masses generally
satisfy the $\Sigma_c$ sum rule, and all predict large positive values for the
$\Sigma_c$ mass combination in Eq.\ (1).

The experimental input  that has been used for this combination of $\Sigma_c$
masses are the two separate mass difference measurements
\begin{eqnarray}
\Sigma_c^{++}-\Sigma_c^0 & = & 0.6\pm .2\quad {\rm Ref.[5]}\\
\Sigma_c^{+}-\Sigma_c^0 & = & 1.4\pm .6\quad{\rm Ref.[12].}
\end{eqnarray}
The $\Sigma_c^{++}-\Sigma_c^0$ mass difference results from four separate
experiments that are reasonably consistent with one another, while there is only
one experiment\cite{crawf} that has measured the $\Sigma_c^+ -\Sigma_c^0$
difference.  There is no reason to question this experimental measurement of
$\Sigma_c^+ -\Sigma_c^0$, and the result of Ref.\ [12] for $\Sigma_c^{++}
-\Sigma_c^0$ agrees well with the other experiments\cite{crawsr}.  However, the
extreme importance of the large discrepancy in the $\Sigma_c$ sum rule of Eq.\
(1) should make a new experimental measure of the mass difference $\Sigma_c^+
-\Sigma_c^0$ a high priority.

The new experimental measurement of the $\Xi_c^\prime$ masses\cite{c98} makes it
possible,
in principle, to test sum rules for separate mass differences of the $\Sigma_c$.
These are
\begin{eqnarray}
\Sigma_c^{++}-\Sigma_c^{0} & = & \Sigma^{*+}-\Sigma^{*-}
+2[(\Xi^{*-}-\Xi^{*0})+(\Xi_c^{\prime +}-\Xi_c^{\prime 0})] \\
(0.6\pm .2) &  & \hspace{.9in} (-6.2 \pm 9.7)\nonumber\\
\Sigma_c^{+}-\Sigma_c^{0} & = & \Sigma^{*0}-\Sigma^{*-}
+(\Xi^{*-}-\Xi^{*0})+(\Xi_c^{\prime +}-\Xi_c^{\prime 0})\\
(1.4\pm .6) &  &\hspace{.9in} (-4.2 \pm 4.9)\nonumber
\end{eqnarray}
Unfortunately, the experimental errors on the $\Xi_c^\prime$ mass differences
are
still too large at this point to make an accurate comparison with
the $\Sigma_c$ mass differences.

Although the discrepancy noted above for the $\Sigma_c$ mass differences puts
any other quark model study of charmed baryons into question, we now look at sum
rules for medium-strong mass differences, anticipating some eventual resolution
(theoretical or experimental) of the difficulties posed by the $\Sigma_c$ mass
splittings.
A new measurement\cite{brand} of the masses of the $\Sigma^{*++}$ and
$\Sigma^{*0}$ baryons makes possible a more accurate test of the sum
rule\cite{cb3}
\begin{eqnarray}
(\Sigma_c^{*+}-\Lambda_c^+)+\frac{1}{2}(\Sigma_c^+ -\Lambda_c^+)
 & = & (\Sigma^{*0}-\Lambda^0)+\frac{1}{2}(\Sigma^0 -\Lambda^0).\\
(319\pm 2)\hspace{.55in} & &\hspace{.55in} (307)\nonumber
\end{eqnarray}
We use the measured $\Sigma_c^{*++}$ mass for the $\Sigma_c^{*+}$ mass,
but that difference is probably small.
A corresponding sum rule\cite{cb3} for the b-quark baryons $\Sigma_b^{*0}$,
$\Sigma_b^{0}$, $\Lambda_b^{0}$ has not changed,
and is in good agreement .

    In Ref.\cite{cb2} we used a sum rule to predict
$2583\pm 3$ MeV for the
$\Xi_c^{\prime +}$ mass.  This mass has now been measured\cite{c98},
and is listed in Table I.  This permits a
test of the sum rule, which we write here as\newpage
\begin{eqnarray}
\Sigma_c^{++}+\Omega_c^{0} -2\Xi_c^{\prime +}
 & = & \Sigma^+ +\Omega^{-} -\Xi^0 -\Xi^{*0}\\
(10\pm 8)\hspace{.38in} & &\hspace{.44in} (15)\nonumber
\end{eqnarray}
The two sum rules in Eqs.\ (6) and (7) are satisfied to about the same
extent as light-quark baryon sum rules relating spin-$\frac{1}{2}$ baryon masses
to spin-$\frac{3}{2}$baryon masses.\cite{cb,sqm}

The new experimental measurements can be used to improve the accuracy
of our previous prediction[3] of the
as yet unmeasured $\Omega_c^{*0}$ mass
\begin{equation}
\Omega_c^{*0}=\Omega_c^{0}+2(\Xi_c^{*+}-\Xi_c^{\prime+})
-(\Sigma_c^{*++}-\Sigma_c^{++})=2779\pm9,
\end{equation}

In conclusion, we can say that increasingly accurate experimental mass
determinations
are making the model independent sum rules discussed here increasingly
useful tests of the quark model for charmed baryons.
We see that sum rules for medium-strong energy differences are satisfied
 at least as well for heavy-quark baryons as for
 light-quark baryons.  However there remains a serious disagreement
 for the $\Sigma_c$  isospin breaking sum rule,
which is violated by three standard deviations.  Since sum rules
in disagreement are of more concern than those which are satisfied,
resolving the $\Sigma_c$  mass differences is of prime importance.
Thus far no theoretical suggestion has been forthcoming.


\begin{thebibliography}{8}
\bibitem[15]{crawsr}Reference \cite{crawf} is the only experiment to
simultaneously measure both
$\Sigma_c$ mass differences.  For this single experiment, the $\Sigma_c$ sum
rule combination gives
$\Sigma_c^{++}+\Sigma_c^{0}-2\Sigma_c^{+}=-1.7\pm 1.0$.
\end{thebibliography}
\end{document}